\documentclass[preprint,showpacs,showkeys,preprintnumbers,amsmath,amssymb,superscriptaddress]{revtex4}
\usepackage{graphicx}
\usepackage{dcolumn}
\usepackage{bm}

\begin{document}

\title{A Comparison of Remnants in Noncommutative Bardeen Black Holes}

\author{S. Hamid Mehdipour}

\email{mehdipour@liau.ac.ir}

\affiliation{Department of Physics, College of Basic Sciences,
Lahijan Branch, Islamic Azad University, P. O. Box 1616, Lahijan,
Iran}

\author{M. H. Ahmadi}

\email{ahmadi@liau.ac.ir}

\affiliation{Department of Physics, College of Basic Sciences,
Lahijan Branch, Islamic Azad University, P. O. Box 1616, Lahijan,
Iran}

\date{\today}

\begin{abstract}
We derive the mass term of the Bardeen metric in the presence of a
noncommutative geometry induced minimal length. In this setup, the
proposal of a stable black hole remnant as a candidate to store
information is confirmed. We consider the possibility of having an
extremal configuration with one degenerate event horizon and compare
different sizes of black hole remnants. As a result, once the
magnetic charge $g$ of the noncommutative Bardeen solution becomes
larger, both the minimal nonzero mass $M_0$ and the minimal nonzero
horizon radius $r_0$ get larger. This means, subsequently, under the
condition of an adequate amount of $g$, the three parameters $g$,
$M_0$, and $r_0$ are in a connection with each other linearly.
According to our results, a noncommutative Bardeen black hole is
colder than the noncommutative Schwarzschild black hole and its
remnant is bigger, so the minimum required energy for the formation
of such a black hole at particle colliders will be larger. We also
find a closely similar result for the Hayward solution.
\end{abstract}


\maketitle

\section{\label{sec:1}Introduction}
The issue of central singularity of a Black Hole (BH) is an open
problem in BH physics. Although it is commonly accepted that only a
not yet accessible quantum theory of gravity would be competent to
solve the problem appropriately, several phenomenological scenarios
have been considered in the literature in order to study BHs with
regular centers (for a review, see \cite{ans}). In 1968, Bardeen
\cite{bar} introduced a compact object with an event horizon and
without an intrinsic singularity, namely Bardeen BH; it is the first
regular BH model in general relativity. The Bardeen spacetime is
spherically symmetric without violating the weak energy condition
and the inside of the horizon is deSitter-like wherein the matter
has a high pressure. In 2000, Ay\'{o}n-Beato and Garc\'{\i}a
\cite{bea} reinterpreted the Bardeen model as the gravitational
field of a nonlinear magnetic monopole. A few years later, Hayward
\cite{hay} investigated the formation and evaporation of a new kind
of the regular solution, i.e. Hayward BH, in which its static region
is Bardeen-like and the dynamic regions are Vaidya-like. Recently, a
family of rotating regular solutions have been obtained by applying
the Newman-Janis algorithm to the Hayward and to the Bardeen
spacetimes \cite{bam}. Afterwards, a general class of regular
solutions using a general mass term described by a function which
generalizes the Bardeen and Hayward mass terms have been constructed
\cite{nev}. The regular BHs have extensively been studied in the
recent literature (see for instance,
\cite{azr1,azr2,azr3,azr4,ghosh,garc,tosh,ghosh2,stuc}).

Besides, the authors of Refs.~\cite{nic1,nic2,nic3,nic4,nic5} have
utilized a spherically symmetric matter distribution leading to no
curvature singularity. In fact, they have presented a physically
inspired type of noncommutativity corrections to BH solutions. In
this method, the point-like structure of mass, instead of being
totally localized at a point, is described by a smeared structure
throughout a region of linear size $\sqrt{\theta}$. In other words,
the mass density of a static, spherically symmetric, particle-like
gravitational source cannot be a delta function distribution, but
will be given by a Gaussian distribution
$\rho_{\theta}(r)=M/(4\pi\theta)^{3/2}\exp(-r^2/4\theta)$. It has
been demonstrated that the modified metric does not allow the BH to
decay below the Planckian relic. The evaporation process ends when
the size of the BH reaches a Planck-sized remnant, explaining the BH
released from the curvature singularity at the origin. Here, the
regularity of the metrics emerges from the appearance of a minimal
length preparing a natural cut-off at small scales. The idea of a
minimal length is confirmed by many outcomes of various approaches
to quantum gravity \cite{sny1,sny2,sny3,sny4,sny5}. This universal
cut-off is entered into the energy-momentum tensor of the Einstein
equations, and stands for the degree of delocalization of the matter
distribution \cite{sma1,sma2,sma3}.

From the other viewpoint, all the various arguments concerning the
so-called Hawking information loss paradox rely on semi-classical
methods and guess about the behavior of systems in the quantum
gravitational regime, but there are substantial struggles over the
success of the arguments. The basis of the information loss problem
turns back to Hawking's discovery that the theory of quantum fields
in a curved spacetime indicates that BHs will emit thermal radiation
at a temperature inversely proportional to their mass \cite{haw}.
Conservation of energy points out that the BH will lose mass through
this procedure, and if nothing stops the evaporation of the BH will
ultimately cease to exist. This proposition of total evaporation is
necessary for Hawking's argument, and may be refused by a remnant
proposal that we will be considering in this paper (for reviews on
resolving the paradox, see
\cite{pre1,pre2,pre3,pre4,chen,dymn,bank}). In other words, Planck
scale physics may terminate the Hawking radiation and prohibit the
appearance of a singularity in the center of a BH, i.e. the
appearance of a BH remnant; for example, in a recent result by Paul
and Majhi \cite{pau}, the nature of the cascade of Hawking emission
spectrum in the presence of a back reaction was studied. They
observed that under a physical background, below a particular value
of the mass, which is of the Plank mass order, the Hawking radiation
must stop wherein a remnant is formed.

In this work we shall include the influences of inspired
noncommutativity to the one of the most popular models of
non-singular BHs, i.e. the Bardeen BH, and analyze the remnants of
Noncommutative Bardeen (NB) BHs. Afterwards, we consider the
thermodynamic properties of the NB solution, providing its Hawking
temperature. Throughout this paper natural units are used with the
following definitions: $\hbar = c = G = k = 1 $.

\section{\label{sec:2}The NB solution}
\begin{figure}[htp]
\begin{center}
\includegraphics{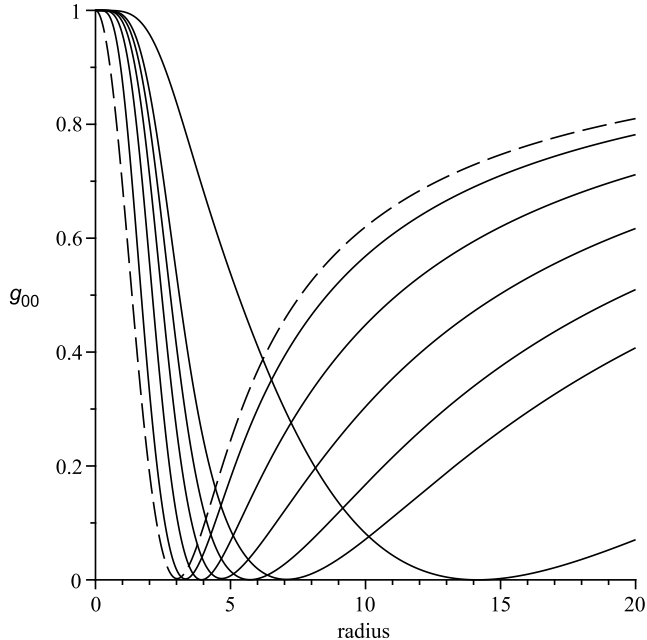}
\end{center}
\vspace{7.5 cm} \caption{\scriptsize {The temporal component of the
metric, $g_{00}$, versus the radius $r/\sqrt{\theta}$ for different
values of $g/\sqrt{\theta}$. The figure shows the possibility of
having an extremal configuration with one degenerate event horizon
at $M=M_0$ (extremal NBBH). This shows the existence of a minimal
nonzero mass ($M_0$) that the BH can shrink to. On the right-hand
side of the figure, from top to bottom, the solid lines correspond
to the NBBH for $g=1.00\sqrt{\theta},~ 2.00\sqrt{\theta},~
3.00\sqrt{\theta},~ 4.00\sqrt{\theta},~ 5.00\sqrt{\theta},$ and
$g=10.00\sqrt{\theta}$, respectively. The dashed line refers to the
Schwarzschild case so that it corresponds to $g=0$.}} \label{fig:1}
\end{figure}

In this section we first include the noncommutative effects in the
line element of Bardeen and then analyze the consequences of the
resulting metric. Now, let us start from the Schwarzschild-like
class of metrics which describe the spacetimes in the so-called
Kerr-Schild form and in the presence of matter
\begin{equation}
\label{mat:1}ds^2=ds_M^2-\frac{f(r)}{r^2}\left(k_\mu
dx^\mu\right)^2,
\end{equation}
where $ds_M^2$ is the Minkowski line element expressed in a
spherical basis and $k_\mu$ is a null vector in Minkowski
coordinates. The expression $f(r)$ is found to be
\begin{equation}
\label{mat:2}f(r)=2m(r)r.
\end{equation}
According to the Kerr-Schild decomposition, the above equation has a
general validity, so its generic structure is kept and it is not
sensitive to different forms of the mass term $m(r)$. For the
Bardeen metric we have
\begin{equation}
\label{mat:3}m(r)=M\left(\frac{r^2}{r^2+g^2}\right)^{\frac{3}{2}},
\end{equation}
where $g$ is the magnetic charge of the BH. Now, using the
noncommutativity approach \cite{nic1,nic2,nic3,nic4,nic5}, the
metric describing the noncommutative geometry inspired Bardeen BH is
given by
\begin{equation}
\label{mat:4}ds^2=\left(1-\frac{2m(r)}{r}\right)dt^2-
\left(1-\frac{2m(r)}{r}\right)^{-1}dr^2-r^2 d\Omega^2,
\end{equation}
where $m(r)$ can now be written in terms of the smeared mass
distribution $M_\theta$ as follows:
\begin{equation}
\label{mat:5}m(r)=M_\theta\left(\frac{r^2}{r^2+g^2}\right)^{\frac{3}{2}},
\end{equation}
and $M_\theta$ can implicity be given in terms of the lower
incomplete Gamma function,
\begin{equation}
\label{mat:6}M_\theta=\int_0^r\rho_{\theta}(r)4\pi
r^2dr=\frac{2M}{\sqrt{\pi}}\gamma\bigg(\frac{3}{2};\frac{r^2}{4\theta}\bigg).
\end{equation}
The radiating behavior of such a noncommutative regular BH can now
be easily investigated by plotting the temporal component of the
metric, $g_{00}$, versus the radius $r$ for an extremal BH with
different values of $g$ (see Fig.~(\ref{fig:1})). The plot presented
in Fig.~\ref{fig:1} shows, for several values of minimal nonzero
mass $M_0$, the possibility of having an extremal configuration with
one degenerate event horizon as the parameter $g$ grows. As this
figure shows, the coordinate noncommutativity leads to the existence
of a remnant mass in which the NBBH can shrink to.

\begin{figure}[htp]
\begin{center}
\includegraphics{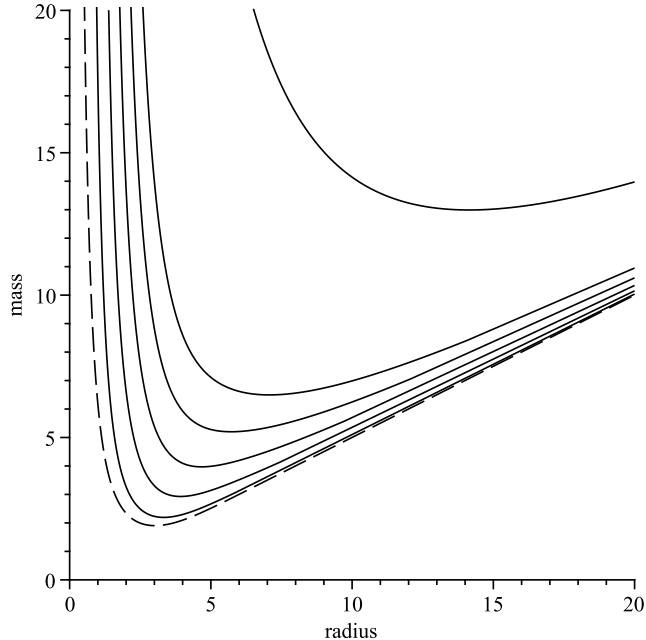}
\end{center}
\vspace{7.5 cm} \caption{\scriptsize {The mass of the NBBH,
$M/\sqrt{\theta}$, versus the event horizon radius,
$r_H/\sqrt{\theta}$, for different values of $g/\sqrt{\theta}$. On
the right-hand side of the figure, from bottom to top, the solid
lines correspond to the NBBH for $g=1.00\sqrt{\theta},~
2.00\sqrt{\theta},~ 3.00\sqrt{\theta},~ 4.00\sqrt{\theta},~
5.00\sqrt{\theta},$ and $g=10.00\sqrt{\theta}$, respectively. The
dashed line refers to the Schwarzschild case so that it corresponds
to $g=0$. }} \label{fig:2}
\end{figure}

The line element (\ref{mat:4}) has a coordinate singularity at the
event horizon radius, $r_H$, that can be obtained from the equation
$g_{00}(r_H)=0$ as follows
\begin{equation}
\label{mat:7}1 - \frac{2m(r_H)}{r_H}=0,
\end{equation}
with
\begin{equation}
\label{mat:8}
m(r_H)=M_\theta(r_H)\left(\frac{r_H^2}{r_H^2+g^2}\right)^{\frac{3}{2}}.
\end{equation}
The analytical solution of Eq.~(\ref{mat:7}) for $r_{H}$ in a closed
form is impossible, but it is possible to solve it to find $M$,
which provides the mass of the NBBH as a function of the horizon
radius $r_H$. This leads to
\begin{equation}
\label{mat:9}
M=\frac{r_H}{2\left(\frac{r_H^2}{r_H^2+g^2}\right)^{\frac{3}{2}}\left[{\cal{E}}
\left(\frac{r_H}{2\sqrt{\theta}}\right)-\frac{r_H}{\sqrt{\pi\theta}}
e^{-\frac{r_H^2}{4\theta}}\right]},
\end{equation}
where the Gauss error function ${\cal{E}}(x)$ is defined by
${\cal{E}}(x)\equiv 2/\sqrt{\pi}\int_{0}^{x}e^{-p^2}dp$. The results
of the numerical solution of the mass as a function of the horizon
radius are displayed in Fig.~\ref{fig:2}. As expected, from the mass
equation (\ref{mat:9}), the noncommutativity indicates a minimal
nonzero mass in order to have an event horizon. So, in the
noncommutative case, for $M < M_0$ there is no event horizon.

For more details, the numerical results for the remnant size of the
BH and for different values of $g/\sqrt{\theta}$ are presented in
Table~\ref{tab:1} which are comparable to Fig.~\ref{fig:2}.
According to Table~\ref{tab:1}, as $g$ increases both the minimal
nonzero mass and the minimal nonzero horizon radius are enlarged
which subsequently lead us to the final result:
\begin{equation}
\label{mat:10}g\propto M_0\propto r_0.
\end{equation}
This means in the limit $g/\sqrt{\sigma}\gg1$, the magnetic charge
is proportional to the remnant mass and to the remnant radius.

\begin{table}
\caption{\scriptsize{The minimal nonzero mass of the NBBH (remnant
mass, $M_0/\sqrt{\theta}$) and also the minimal nonzero horizon
radius, $r_0/\sqrt{\theta}$, for different values of
$g/\sqrt{\theta}$. For a large amount of $g/\sqrt{\theta}$, i.e.
$g/\sqrt{\theta}\gg1$, there is a linear relationship between the
remnant mass and the remnant radius. As can be seen from the table,
the results are comparable to Fig.~\ref{fig:2}.}}
\begin{center}
\begin{tabular}{c|c|c}
\hline \hline
\multicolumn{3}{c}{NBBH} \\
\hline  Magnetic Charge &   Remnant Mass & Remnant Radius  \\
\hline$g=0$ & $M_0\approx1.90\sqrt{\theta}$ & $r_0\approx3.02\sqrt{\theta}$\\
\hline$g=1.00\sqrt{\theta}$ & $M_0\approx2.19\sqrt{\theta}$ & $r_0\approx3.33\sqrt{\theta}$ \\
\hline$g=2.00\sqrt{\theta}$ & $M_0\approx2.92\sqrt{\theta}$ & $r_0\approx3.92\sqrt{\theta}$ \\
\hline$g=3.00\sqrt{\theta}$ & $M_0\approx3.96\sqrt{\theta}$ & $r_0\approx4.67\sqrt{\theta}$\\
\hline$g=4.00\sqrt{\theta}$ & $M_0\approx5.20\sqrt{\theta}$ & $r_0\approx5.72\sqrt{\theta}$\\
\hline$g=5.00\sqrt{\theta}$ & $M_0\approx6.49\sqrt{\theta}$ & $r_0\approx7.07\sqrt{\theta}$\\
\hline$g=10.00\sqrt{\theta}$ & $M_0\approx12.99\sqrt{\theta}$ & $r_0\approx14.14\sqrt{\theta}$\\
\hline$g=100.00\sqrt{\theta}$ & $M_0\approx129.90\sqrt{\theta}$ & $r_0\approx141.42\sqrt{\theta}$\\
\vdots & \vdots & \vdots \\
\hline \multicolumn{3}{c}{$g\gg\sqrt{\theta}$ \qquad
$\Longrightarrow$
\qquad $g\propto M_0\propto r_0$} \\
\hline \hline
\end{tabular}
\end{center}
\label{tab:1}
\end{table}

Here, it should be emphasized that the physical interpretation of
the noncommutative parameter $\theta$ is the smallest fundamental
cell of an observable area in noncommutative geometry, in the same
way that the Planck constant $\hbar$ explains the smallest
fundamental cell of an observable phase space in quantum mechanics.
The scale of $\sqrt{\theta}$ is, possibly and most reasonably, of
the order of an inverse characteristic energy of the Planck scale.
Most of the phenomenological examinations of the noncommutativity
models have assumed that the noncommutative energy scale cannot lie
far above the TeV regime \cite{hin,riz1,riz2,riz3}. Given that the
fundamental Planck scale in models with large extra dimensions is as
small as a TeV, for solving the hierarchy problem
\cite{ant1,ant2,ant3,ant4}, therefore it could be possible to set
the noncommutative effects in a TeV energy scale.

In this process, the minimum value of the NBBH mass increases to a
value more than its value for the noncommutative Schwarzschild one
(see Table~\ref{tab:1}). Thus, in the theory of regular BHs, if the
parameter $g$ becomes sufficiently large, i.e. $g/\sqrt{\theta}\gg1$
with a sufficiently small noncommutative inverse length parameter
($1/\sqrt{\theta}\sim 1$ TeV), then the noncommutativity effect can
concretely decrease the possible formation and detection of BHs in
TeV-scale collisions at particle colliders, such as the Large Hadron
Collider (LHC) \footnote{Note that, a micro BH can be produced at
the LHC just under the condition of $E_{cm} > M_0$, where $E_{cm}$
is the parton-parton center-of-mass energy which is proportional to
the TeV energy scale.} and the Ultra-High Energy Cosmic Ray (UHECR).
However, if the fundamental Planck scale is of the order of a few
TeV, then the LHC may produce BHs. These BHs may have masses on the
order of TeV. In this sense, on the other hand, the complete decay
of BHs is impossible and the final Planck-sized remnant can be
thought of as the order of TeV.

Based on our computations, the total evaporation of the BH is not
possible in principle. Therefore, the idea of a stable BH remnant as
a candidate to conserve information has fixed. Note that, currently
there are some proposals about what happens to the information that
falls into a BH. One of the main proposals is that the BH never
disappears completely, and the information is not lost, but would be
stored in a stable remnant. A remnant proposal reflects the fact
that the semi-classical approaches used to derive the Hawking effect
are obviously inapplicable when the BH reaches the Planck mass.
Possibly, Planckian physics will propose some way of conserving the
information held in the BH.

The remnant proposals are generally divided into two categories. The
first is stable remnants and the second is long-lived remnants. If a
remnant is stable, then Planck scale quantum gravitational effects
shut down the Hawking radiation, and the BH remnant continues to
exist for all future time. Inasmuch as the BH, and its information,
are prohibited from vanishing totally, the troubling result of
Hawking's argument, i.e. the non-unitary evolution is prevented. For
example, in Ref.~\cite{meh2}, we have shown that, as a well-known
result of the spacetime noncommutativity, a part of information may
be preserved in a stable BH remnant. On the other hand, the process
for long-lived remnants is completely different. They eventually
disappear. In spite of the fact that the semi-classical models imply
that no information can escape in the Hawking radiation, these
models will fail in Planckian processes. This scenario is inspired
by this cognition that the physics at the Planck scale might return
the information to the external universe, and once the coherence of
the external universe is made safe, the remnant could safely
disappear. As an example of long-lived kinds of remnants, one can
point to Ref.~\cite{meh} in which an exact $(t - r)$ dependent case
of a noncommutative Schwarzschild-like metric for a Vaidya solution
was calculated. As an important result of Ref.~\cite{meh}, the idea
of a stable BH remnant as a candidate to conserve information has
been failed which means that if we pick up a time-dependent Gaussian
distribution of mass/energy, then it will be possible to find a zero
remnant mass, albeit in a long-time limit.

One of the serious problems with these remnants is the probability
of their detection. Given that the interactions of BH remnants are
purely gravitational, the cross-section is highly small, and a
direct observation of these remnants seems impossible. A possible
indirect evidence might be related to the cosmic gravitational wave
background. In contrast to photons, the gravitons radiated during
evaporation would be instantly frozen. Since, the BH evaporation
finishes when it reduces to a remnant, hence, the graviton spectrum
should have a cut-off at the Planck mass. In general, such a cut-off
is expected to have a redshift on the order of $10^{14}$ GeV.
Moreover, we know that the nature of dark matter is hitherto
remained an open problem. There exist many dark matter candidates in
which most of them are non-baryonic weakly interacting massive
particles. A candidate which is not closely connected to particle
physics is the relics of primordial BHs \cite{zel}. Some specific
inflation models naturally induce a great number of such BHs, e.g.
the hybrid inflation model can generally produce a required
abundance of primordial BH remnants for them to be the main source
of the dark matter \cite{lind}.

\section{\label{sec:3}The NBBH temperature}
While we do not yet have any credible candidate for a full quantum
gravity theory, more phenomenological procedures have tried to
investigate micro BHs. The recent anticipations imply the
conceivable results of LHC experiments, containing the creation of
these objects. The possible experimental production of BHs at
particle colliders is one of the most significant subfields in extra
dimension models. These newly formed miniature BHs first lose their
hairs associated with the multipole and the angular momenta, then
classically reach the stable Schwarzschild solutions, and finally
evaporate via Hawking radiation up to promising Planck-sized
remnants. The Hawking temperature is generally subjected to
corrections from many sources, particularly, those related to a BH
with the mass of the order of the Planck mass. Hence, the study of
TeV-scale BHs in the UHECR and particle colliders requires a perfect
examination of how temperature corrections affect BH thermodynamics.

In the following, with the above motivation, we would like to find
the temperature corrections of the NB solution. When the NBBH
radiates, its Hawking temperature can be calculated to find
\begin{eqnarray}
\label{mat:11}T_H=\frac{1}{4\pi}\frac{dg_{00}}{dr}\bigg|_{r=r_H}=\frac{M}{4\sqrt{(\pi\theta)^3}\left(r_H^2+g^2\right)
^{\frac{5}{2}}} \bigg[4r_H\sqrt{\pi\theta^3}
\left(\frac{r_H^2}{2}-g^2\right){\cal{E}}\left(\frac{r_H}{2\sqrt{\theta}}\right)\nonumber
\\-r_H^2e^{-\frac{r_H^2}{4\theta}}\left(r_H^4+2r_H^2\theta+r_H^2g^2-4\theta
g^2\right)\bigg].
\end{eqnarray}
For large BHs, $r_H/\sqrt{\theta}\gg1$, and $g=0$, the Gauss error
function tends to unity and the exponential term is reduced to zero.
Thus, one recovers the standard result for the temperature of a
Schwarzschild BH, i.e. $T_H=M/(2\pi r_H^2)=1/(4\pi r_H)$.

\begin{figure}[htp]
\begin{center}
\includegraphics{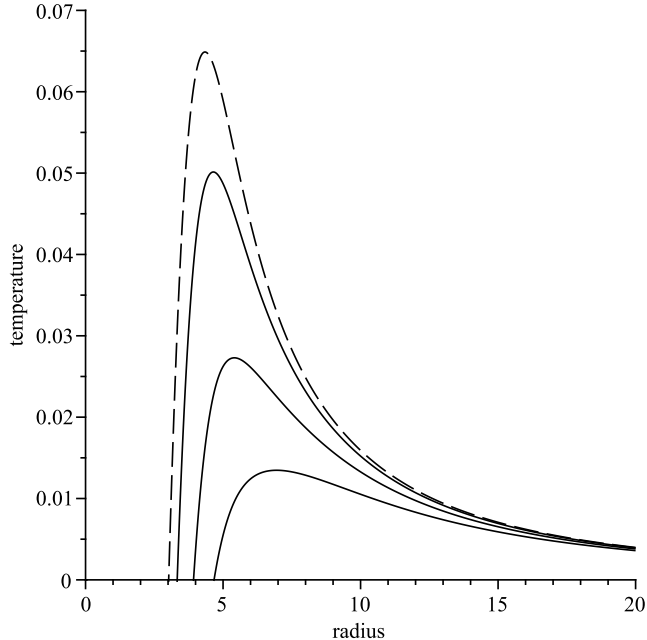}
\end{center}
\vspace{7.5 cm} \caption{\scriptsize {The temperature $T_H$ versus
the horizon radius, $r_H/\sqrt{\theta}$. We have set
$M=10.00\sqrt{\theta}$. On the right-hand side of the figure, from
top to bottom, the solid lines correspond to the NBBH for
$g=1.00\sqrt{\theta}$, $g=2.00\sqrt{\theta}$, and
$g=3.00\sqrt{\theta}$, respectively. The dashed line refers to the
Schwarzschild case so that it corresponds to $g=0$.}} \label{fig:3}
\end{figure}

At this stage, the numerical result of the Hawking temperature
versus the horizon radius is presented in Fig.~\ref{fig:3}. This
figure shows that the temperature peak drops with increasing the
parameter $g$. Therefore, we expect that the NBBHs to be colder than
the noncommutative Schwarzschild BHs. In addition, the size and the
mass of the NBBH remnant at the final stage of the evaporation
increase with increasing the magnetic charge. Hence, the remnant of
a NBBH may be big compared to the noncommutative Schwarzschild case.

\begin{figure}[htp]
\begin{center}
\includegraphics{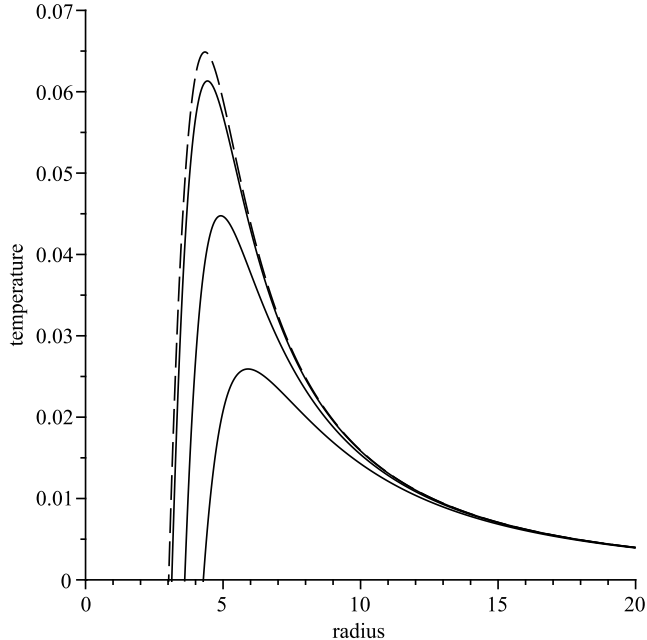}
\end{center}
\vspace{7.5 cm} \caption{\scriptsize {The temperature $T_H$ versus
the horizon radius, $r_H/\sqrt{\theta}$. We have set
$M=10.00\sqrt{\theta}$. On the right-hand side of the figure, from
top to bottom, the solid lines correspond to the NHBH for
$g'=1.00\sqrt{\theta}$, $g'=2.00\sqrt{\theta}$ and
$g'=3.00\sqrt{\theta}$, respectively. The dashed line refers to the
Schwarzschild case so that it corresponds to $g'=0$.}} \label{fig:4}
\end{figure}

As an important note, if we had chosen the Hayward solution, as
another popular example of regular BHs, solely the mass term would
have altered, however the general properties would have directed to
entirely comparable consequences to those above. Now, let us
consider the metric describing the Noncommutative Hayward (NH) BH,
in Schwarzschild coordinates, which is immediately given by
Eq.~(\ref{mat:4}) with this new mass term
\begin{equation}
\label{mat:12}m(r)=M_\theta\left(\frac{r^3}{r^3+g'^3}\right),
\end{equation}
where $g'$ is a positive constant measuring the deviations from the
classical Kerr metric. The lack of responsiveness of the results to
the kind of the regular BH can be easily exhibited by plotting the
Hawking temperature as a function of radius for NHBHs (see
Fig.~\ref{fig:4}). Comparing these results with the results of
Fig.~\ref{fig:3} shows the close similarity of outcomes in these two
types of regular BHs.

At this point it is worth pointing out that most of the results in
the noncommutative framework are confirmed by the so-called
Generalized Uncertainty Principle (GUP) context
\cite{gup10,gup11,gup12,gup13,gup14}. It is widely accepted that the
Heisenberg uncertainty principle should be reformulated owing to the
noncommutative nature of spacetime at the Planck scale. The
application of the GUP to BH thermodynamics has attracted
considerable attention in the literature which leads to significant
modifications to the emission process, particularly at the final
stages of the BH evaporation (there is a large body of literature on
this subject; see for example, \cite{gup20,gup21,gup22,gup23}). As a
result of GUP effects on this issue, we have shown that a
modification of the de Broglie relation and corresponding
commutation relations in the quantum tunneling framework of the BH
evaporation lead to correlations between emitted modes of
evaporation \cite{meh1,meh11,meh12}. In this setup, information
leaks out of the BH in the form of non-thermal GUP correlations and,
on the other hand, the inclusion of quantum gravity effects as the
GUP expression can halt the evaporation process, so that a stable BH
remnant is left behind, including a part of the BH information
content (see also \cite{meh2}). In addition, recently, the authors
of Ref.~\cite{fen} investigated the GUP effect on the thermodynamics
of a Schwarzschild-Tangherlini BH and found that the GUP corrected
Hawking temperature is smaller than the original case; it goes to
zero when the mass of the BH reaches a minimal value, which is
supported by the results obtained in the framework of the inspired
noncommutativity. Also, the results of their study concerning the
possibilities to observe a micro BH in the LHC have shown that the
minimum energy for the production of the BH is larger than the
current energy scales of LHC.

As a final remark, our results are supported by the results obtained
in the framework of gravity's rainbow \cite{ali1,ali2,ali3}. Ali and
his coworkers \cite{ali1,ali2,ali3} have argued that since a remnant
depends critically on the structure of the rainbow functions,
therefore a remnant is formed for all black objects in the context
of gravity's rainbow and this is a model-independent phenomenon.
Their calculations have shown that the behavior of Hawking's
radiation changes considerably near the Planck scale in gravity's
rainbow such that black objects do not evaporate completely and a
remnant is left. Moreover, they have found that the mass of their
remnant is greater than the energy scale at which experiments were
carried out at the LHC \cite{ali4}.

\section{\label{sec:4}Summary}
In summary, we have applied the noncommutativity effects to Bardeen
BHs. The noncommutative effects become susceptible when the mass of
the BH reaches the order of the Planck scale, it stops radiating and
leads to a BH remnant. It is concluded that, for an adequately large
magnetic charge of a NBBH there is a linear relationship between the
remnant mass and the remnant radius that is just the same as
appeared in the relation between the horizon radius and the BH mass
for the standard Schwarzschild case. We have found that the
temperature peak of the NBBH decreases as the parameter $g$
increases. Thus, a NBBH is colder than a noncommutative
Schwarzschild BH. In this setup, the final stage of the evaporation
of a noncommutative regular BH is a remnant in which it has an
increasing size with raising its own characteristic parameter. As a
consequence, in the theory of noncommutative regular BHs, the
minimum value of energy for the production of such a BH at the
current energy scales of LHC is larger, so the possibility for its
detection is less.

\section*{Acknowledgments}
Financial support by Lahijan Branch, Islamic Azad University Grant
No. 17.20.5.3517 is gratefully acknowledged. The authors thank to C. Bambi for valuable suggestions.\\

\end{document}